\documentclass[%
 reprint,showkeys,showpacs,
 amsmath,amssymb,
 aps,prd,
]{revtex4-1}
\usepackage{hyperref}
\begin{document}
\title[Cosmological Bonnor-Melvin]{Bonnor-Melvin universe with a cosmological constant}
\author{Martin \v{Z}ofka}
\email{zofka@mbox.troja.mff.cuni.cz}
\affiliation{Institute of Theoretical Physics, Faculty of Mathematics and Physics, Charles University, Czech Republic}
\vspace{10pt}
\begin{abstract}
We generalize the well-known Bonnor-Melvin solution of the Einstein-Maxwell equations to the case of a non-vanishing cosmological constant. The spacetime is again cylindrically symmetric and static but, unlike the original solution, it truly represents a homogeneous magnetic field.
\end{abstract}
\maketitle
\section{Introduction}
Magnetic fields play an important role in many astrophysical phenomena across a range of distance scales from stars and accretion disks to galactic nuclei and intergalactic regions. As they often occur in the vicinity of compact massive objects or in strong gravitational fields it is important to study them in the context of general relativity as well. One of interesting exact solutions of Einstein-Maxwell equations is the Bonnor-Melvin universe describing a static, cylindrically symmetric (electro)magnetic field immersed in its own gravitational field \cite{Bonnor,Melvin}. The magnetic field is aligned with the symmetry axis. This is one possible analogy of the classical homogeneous magnetic field. However, since the magnetic field contributes to the energy-momentum tensor, which curves the spacetime, the field needs to decrease away from the axis, so as not to collapse on itself, and the scalar invariant, $F_{\mu \nu} F^{\mu \nu}$, is thus not constant unlike in the classical case and the field is not homogeneous.

Already the original paper by Melvin suggested several possible ways of generalizing the spacetime. If we wish to restore the balance for a homogeneous field, we need to incorporate an element countering the collapse of the field. It thus makes sense to search for such a solution with a non-vanishing---and, in fact, positive---cosmological constant. It is of interest then that the observed intergalactic magnetic fields \cite{Tavecchio,Neronov-Vovk} are also considered to be of cosmological origin and thus related to the large-scale structure of the universe \cite{Subramanian}.
We first briefly review the Bonnor-Melvin case to contrast it with the homogeneous cosmological case.
\section{\textbf{Bonnor-Melvin}}
In cylindrical coordinates, one possible form of the metric reads
\begin{equation} \label{Bonnor Melvin}
g_{\mu \nu} = \alpha^{-2}(-\mbox{d}t^2+\mbox{d}z^2) + \alpha^{-5}\mbox{d}r^2 + \alpha r^2 \mbox{d}\varphi^2,
\end{equation}
where
\begin{equation} \label{Definition of alpha}
\alpha = 1 - K^2 r^2,
\end{equation}
and $r \leq 1/K$, with the upper limit corresponding to proper radial infinity.
The constant $K$ determines the strength of the magnetic field the 4-potential of which is
\begin{equation} \label{4-potential}
A = K r^2 \mbox{d}\varphi.
\end{equation}
The Maxwell tensor only has one non-zero component
\begin{equation} \label{Maxwell tensor}
F = 2 K r \; \mbox{d}r \wedge \mbox{d}\varphi,
\end{equation}
and its invariant reads
\begin{equation} \label{Maxwell field invariant}
F_{\mu \nu} F^{\mu \nu} = 8 K^2 \alpha^4,
\end{equation}
which is obviously non-constant. In flat spacetime with a homogeneous magnetic test field along the $z$-direction with $\vec{B} = B \vec{e}_z$, we have $F = B r \: \mbox{d}r \wedge \mbox{d}\varphi$ and $F_{\mu \nu} F^{\mu \nu} = 2 B^2$. We compare this to the Bonnor-Melvin solution on the axis of symmetry, where the field approaches the Minkowski spacetime. This yields an analogy between the two electromagnetic fields with $B=2K$.

The solution is a Kundt class spacetime of algebraic type D that is not asymptotically flat far away from the axis and that was shown to be stable against radial perturbations \cite{Thorne}. It has been generalized for the case of non-linear electrodynamics \cite{Gibbons+Herdeiro} and there are also cylindrically symmetric, magneto-hydrodynamic cosmological models involving generally non-static perfect fluid \cite{Singh+Yadav, Roy+Prakash}.

We now proceed to formulate the general Einstein-Maxwell equations for a cylindrically symmetric spacetime featuring a magnetic field aligned with the axis and a non-vanishing cosmological constant.
\section{\textbf{General Einstein-Maxwell equations}}
Denoting the proper radius by $r$, we may write a static cylindrically symmetric metric as
\begin{equation}\label{general line element}
ds^2 = -\exp A(r) \; dt^2 + dr^2 + \exp B(r) \; dz^2 + \exp C(r) \;d\varphi^2,
\end{equation}
here $t,z \in \! I\hspace{-0.13cm}R,r \in \! I\hspace{-0.13cm}R^+, \varphi \in [0,2\pi)$. Next, we list components of $G_{\mu\nu}-\Lambda g_{\mu\nu}$:
\begin{widetext}
\begin{eqnarray}
  G_{tt}-\Lambda g_{tt} &=& \frac{1}{4} \mbox{e}^A \left( 2B'' + \left( B' \right)^{2} + 2C'' + \left( C' \right) ^{2} + C'B' + 4\Lambda \right),\\
  G_{rr}-\Lambda g_{rr} &=& - \frac{1}{4} A'B' - \frac{1}{4} A'C' - \frac{1}{4} B'C' - \Lambda,\\
  G_{zz}-\Lambda g_{zz} &=& -\frac{1}{4} \mbox{e}^B \left( 2A'' + \left( A' \right)^{2} + 2C'' + \left( C' \right) ^{2} + A'C' + 4\Lambda \right),\\
  G_{\varphi\varphi}-\Lambda g_{\varphi\varphi} &=& -\frac{1}{4} \mbox{e}^C \left( 2A'' + \left( A' \right)^{2} + 2B'' + \left( B' \right) ^{2} + A'B' + 4\Lambda \right),
\end{eqnarray}
\end{widetext}
with primes denoting derivative with respect to the radial coordinate. We assume a purely magnetic field aligned with the axis of symmetry, yielding thus a Maxwell tensor of the form
\begin{equation}\label{Maxwell}
    F = H(r) \: \mbox{d}r \wedge \mbox{d}\varphi.
\end{equation}
The invariant of the field is
\begin{equation}\label{Maxwell invariant}
    F_{\mu\nu}F^{\mu\nu} = 2H^2\mbox{e}^{-C} \equiv 2f^2,
\end{equation}
where we defined a new quantity, $f(r)$, while $\star F_{\mu \nu} F^{\mu \nu} = 0$. The stress-energy tensor writes
\begin{eqnarray}\label{stress-energy tensor}
  T_{tt} &=& \frac{1}{8\pi} \mbox{e}^{A-C} H^2,\\
  T_{rr} &=& \frac{1}{8\pi} \mbox{e}^{-C} H^2,\\
  T_{zz} &=& -\frac{1}{8\pi} \mbox{e}^{B-C} H^2,\\
  T_{\varphi\varphi} &=& \frac{1}{8\pi} H^2.
\end{eqnarray}
Finally, Einstein equations are equivalent to
\begin{eqnarray}
  0&=& 2(B'' \! + \! C'')\!+\!\left(B'\right)^{2}\!+\!\left(C'\right)^{2}\!+\! B'C'\!+\!4\Lambda\!+\!4f^2,\label{Einstein t}\\
  0&=& 2(A'' \! + \! C'')\!+\!\left( A' \right)^{2}\!+\!\left( C' \right)^{2}\!+\!A'C'\!+\!4\Lambda\!+\!4f^2,\label{Einstein z}\\
  0&=& 2(A'' \! + \! B'')  \! +  \! \left( A' \right)^{2}  \! + \!  \left( B' \right) ^{2}  \! + \!  A'B' \!  +  \! 4\Lambda  \! -  \! 4f^2,\label{Einstein phi}\\
  0&=& A'B' + A'C' + B'C' + 4\Lambda - 4f^2.\label{Einstein r}
\end{eqnarray}
Maxwell equations $\sqrt{-g}{F^{\mu\nu}}_{;\nu}=(\sqrt{-g}{F^{\mu\nu}})_{,\nu} = 0$ are identities apart from
\begin{eqnarray}\label{Maxwell equation}
    \sqrt{-g}{F^{\varphi\alpha}}_{;\alpha} & = & (\sqrt{-g} F^{\varphi r})_{,r} = (\mbox{e}^{\frac{A+B-C}{2}} F_{\varphi r})_{,r} = \nonumber\\
    & = & -(\mbox{e}^{\frac{A+B-C}{2}} H)_{,r} = -(\mbox{e}^{\frac{A+B}{2}} f)_{,r} = 0,
\end{eqnarray}
which yields
\begin{equation}\label{Maxwell constraint}
    \mbox{e}^{\frac{A+B}{2}} f = const.
\end{equation}
However, (\ref{Maxwell constraint}) is a consequence of Einstein equations which can be seen as follows: differentiate (\ref{Einstein r}) and subtract from it $A'$.(\ref{Einstein t}) + $B'$.(\ref{Einstein z}) + $C'$.(\ref{Einstein phi}) to obtain
\begin{eqnarray}
    &&16ff' +  4f^2(A+B-C)' \nonumber\\
    &&+(A+B+C)'(4\Lambda + A'B' + A'C' + B'C') =0,
\end{eqnarray}
where we substitute for the second bracket in the middle term from (\ref{Einstein r}) to yield
\begin{equation}
    2f' + f(A+B)' = 0,
\end{equation}
which can be integrated to yield (\ref{Maxwell constraint}).
\section{The homogeneous solution}
We now specialize to a homogeneous magnetic field, which means that we require the invariant of the field, $F_{\mu\nu}F^{\mu\nu}$, to be constant throughout the spacetime and thus
\begin{equation}\label{Invariant}
    f = const.
\end{equation}
It then follows immediately from (\ref{Maxwell constraint}) that $A+B = const.$ However, (\ref{Einstein phi})+(\ref{Einstein r}) then imply
\begin{eqnarray}\label{fixing f}
    &2(A + B)'' + \left( (A + B)' \right)^{2} + C'(A + B)' + 8(\Lambda - f^2) = 0& \nonumber\\
    &\Rightarrow f^2 = \Lambda.&
\end{eqnarray}
We also note that we thus have $\Lambda>0$, as expected. Equation (\ref{Einstein r}) then immediately shows that $A'B'=0$ while $A'+B'=0$, which yields
\begin{equation}\label{AB}
    A = const., B = const.
\end{equation}
We can always rescale $t$ and $z$ to achieve $A = 0$, $B = 0$. There is only a single Einstein equation remaining, namely
\begin{equation}\label{Sigma}
    C'' + \frac{1}{2} C'^2 + 4\Lambda =0.
\end{equation}
It thus follows that
\begin{equation}
C(r) = 2 \ln \sigma + 2 \ln \sin \left(\sqrt{2 \Lambda} (r+R) \right),
\end{equation}
with $\sigma$ and $R$ integration constants. We shift the radial coordinate, removing thus $R$, to finally  express the metric as
\begin{equation}\label{metric with deficit angle}
ds^2 = -dt^2 + dr^2 + dz^2 + \sigma^2 \sin^2 \left(\sqrt{2 \Lambda} r \right) \;d\varphi^2,
\end{equation}
while the magnetic field reads
\begin{equation}
H(r) = \sqrt{\Lambda} \: \sigma \sin \left(\sqrt{2 \Lambda} r \right).
\end{equation}
As we approach $\sqrt{2 \Lambda} r = \pi$ the circumference of the rings $r=$ const. vanishes which suggests this is the location of an axis of some sorts. We thus now rescale both $r$ and $\varphi$ to bring the line element into the form
\begin{equation}\label{Plebanski-Hacyan}
ds^2 = -dt^2 + dz^2 + \frac{1}{2 \Lambda} \left( dr^2 + \sin^2 \! r \;d\varphi^2 \right),
\end{equation}
with
\begin{equation}
H(r) = \frac{1}{\sqrt{2}} \sin r.
\end{equation}
This is locally a direct product of 2D Minkowski and a 2-sphere of constant radius $1/\sqrt{2 \Lambda}$. The curvature scalars are $\Phi_{11} = \frac{1}{2} \Lambda >0$ and $\Psi_2 = -\frac{1}{3} \Lambda$, and thus they satisfy the condition $2\Phi_{11} + 3\Psi_2 = 0$. Hence, these space-times belong to the ``exceptional electrovacuum type D metrics with cosmological constant'' investigated by Pleba\'{n}ski and Hacyan \cite{Plebanski-Hacyan}, see also \cite{Griffiths+Podolsky}. They admit a six-dimensional group of isometries $ISO(1, 1) \times SO(3)$.

Generally the solution admits a deficit angle due to the presence of $\sigma$ in (\ref{metric with deficit angle}) and the axis is thus singular, forming the spacetime's only singularity. Therefore, the spacetime is in fact a direct product of Minkowski and a squashed sphere at every point. The solution again belongs in the Kundt class. Since this is a homogeneous spacetime, both charged and uncharged test particles can remain static anywhere or just sail along the magnetic field direction at a constant speed. As expected, we also find helical paths.

Using the electromagnetic duality, $F\rightarrow\star F, \star F\rightarrow-F$, the Maxwell field can also be converted to a homogeneous electric field parallel to the cylindrical axis with $E_z=-\sqrt{\Lambda}$ and $F_{\mu \nu} F^{\mu \nu} = 2\Lambda$. It is of interest to look at possible shell sources---we identify cylindrical surfaces of the same circumference located symmetrically with respect to $\sqrt{2\Lambda}r=\pi/2$. We keep the region below the upper radius and as we leave it, heading outwards, we reappear at the lower radius, entering the same region again. We thus get an infinitely thin shell and, using the Israel formalism, we find that the 3-current induced on the shell vanishes while the induced 3D stress-energy tensor only has two non-zero components, corresponding to a positive pressure along the axis equal in magnitude to a negative energy density. This is analogous to the shell source producing a cosmic-string spacetime with a surplus angle as opposed to a deficit one. The resulting electromagnetic field is due to sources at infinity along the axis while the gravitational field is due both to the shell and the stress-energy tensor of the electromagnetic field.

To our knowledge, this is the closest general relativistic analogue of the classical homogeneous magnetic field including particularly the motion of test particles. As an extension of this work we intend to study whether the equations (\ref{Einstein t}-\ref{Einstein r}) admit any other solution at all or whether, in fact, the Pleba\'{n}ski-Hacyan spacetime is the only cylindrically symmetric static solution with the cosmological constant and an aligned magnetic field.

\begin{acknowledgments}
The work was supported by Czech Grant Agency, GACR 17-13525S.
\end{acknowledgments}
%
%We will use a different form of the metric
%%
%\begin{equation}\label{line element fractions}
%ds^2 = -dt^2 + dz^2 + \frac{d\rho^2}{2 \Lambda (1-\rho^2)}  + \mbox{e}^{\sigma} \rho^2 \;d\varphi^2,
%\end{equation}
%%
%with
%%
%\begin{equation}
%H(\rho) = \frac{1}{\sqrt{2}} \mbox{e}^{\frac{\sigma}{2}} \frac{\rho}{\sqrt{1-\rho^2}},
%\end{equation}
%%
%or
%%
%\begin{equation}
%A = -\mbox{e}^{\frac{\sigma}{2}} \sqrt{\frac{1-\rho^2}{2}} \mbox{d}\varphi.
%\end{equation}
%%
%Proper distances
%
%The geodesic equations read
%%
%\begin{eqnarray}
%  \ddot{\varphi}+\frac{2\dot{\rho}\dot{\varphi}}{\rho} = 0,&&\\
%  \ddot{\rho}+\frac{\rho \dot{\rho}^2}{1-\rho^2} -2\Lambda (1-\rho^2) \rho \mbox{e}^\sigma \dot{\varphi}^2 = 0,&&\\
%  \ddot{t} = 0,&&\\
%  \ddot{z} = 0.&&
%\end{eqnarray}
%%
%Killing vectors $[\alpha; 0; \beta; \gamma \rho^2]$.
%
%Shell sources
%
%Singularities: The Kretschmann scalar is $R_{\alpha \beta \gamma \delta} R^{\alpha \beta \gamma \delta} = 16 \Lambda^2$ and constant, Chern-Pontryagin  vanishes $\star R_{\alpha \beta \gamma \delta} R^{\alpha \beta \gamma \delta} = 0$, Euler vanishes $\star R\star_{\alpha \beta \gamma \delta} R^{\alpha \beta \gamma \delta} = 0$.

\end{document}